\documentclass[
preprint,
showpacs,preprintnumbers,
bibnotes,
 amsmath,amssymb,
 aps,
 pra,
superscriptaddress,
longbibliography,
]{revtex4-1}

\usepackage{amsmath}
\usepackage{amsfonts}
\usepackage{amssymb}
\usepackage{graphicx}
\usepackage{xspace}
\usepackage{color}
\usepackage{comment}
\usepackage[hidelinks]{hyperref}
\usepackage{multirow}
\usepackage{dcolumn}
\usepackage{bm}
\usepackage{mathtools}

\usepackage{stackengine}
\setstackgap{S}{2pt}
\setstackgap{L}{.7\baselineskip}

\usepackage{comment}

\usepackage[detect-weight=true, detect-family=true]{siunitx}

\graphicspath{ {./fig/} }

\begin{document}

\title{
Polarization control with plasmonic antenna-tips: A universal approach for optical nano-crystallography and vector-field imaging
}

\author{Kyoung-Duck Park}
\affiliation
{Department of Physics, Department of Chemistry, and JILA,\\ University of Colorado, Boulder, CO, 80309, USA}
\affiliation
{Center for Experiments on Quantum Materials,\\ University of Colorado, Boulder, CO, 80309, USA}
\author{Markus B. Raschke}
\affiliation
{Department of Physics, Department of Chemistry, and JILA,\\ University of Colorado, Boulder, CO, 80309, USA}
\affiliation
{Center for Experiments on Quantum Materials,\\ University of Colorado, Boulder, CO, 80309, USA}
\email{markus.raschke@colorado.edu}
\date{\today}

\begin{abstract}

\noindent 
\textbf{Controlling the propagation and polarization vectors in linear and nonlinear optical spectroscopy enables to probe the anisotropy of optical responses providing structural symmetry selective contrast in optical imaging.
Here we present a novel tilted antenna-tip approach to control the optical vector-field by breaking the axial symmetry of the nano-probe in tip-enhanced near-field microscopy.
This gives rise to a localized plasmonic antenna effect with significantly enhanced optical field vectors with control of both \textit{in-plane} and \textit{out-of-plane} components.
We use the resulting vector-field specificity in the symmetry selective nonlinear optical response of second-harmonic generation (SHG) for a generalized approach to optical nano-crystallography and -imaging.
In tip-enhanced SHG imaging of monolayer MoS$_2$ films and single-crystalline ferroelectric YMnO$_3$, we reveal nano-crystallographic details of domain boundaries and domain topology with enhanced sensitivity and nanoscale spatial resolution.
The approach is applicable to any anisotropic linear and nonlinear optical response, and provides for optical nano-crystallographic imaging of molecular or quantum materials.
} 
\end{abstract}

\maketitle

\noindent Symmetry selective optical imaging of e.g. crystallinity, molecular orientation, and static or dynamic ferroic order and polarization is desirable, yet as of today access to these internal material propeties on the micro- to nano-scale has not been provided in optical microscopy in a systematic way.
Molecular vibrations, phonons, excitons, and spins in their interaction with light give rise to an anisotropic linear and nonlinear optical response.
This optical response is sensitive to the direction of the wavevector and the polarization of the optical driving fields and correlated with the structural symmetries of the material.
In reflection or transmission measurements of far-field optical imaging and spectroscopy, the transverse projection of the optical field as determined by the laws of linear and nonlinear reflection and refraction gives access to the optical selection rules associated with the materials symmetries \cite{anastassakis1997, fiebig2000, najafov2010, yin2014}, yet with limited degrees of freedom constrained by the wavevector conservation in far-field optics.

In contrast, wavevector conservation is lifted in near-field scattering depending on structure and orientation of the nano-objects as scattering element.
In combination with near-field imaging based on tip-scattering, in scanning near-field microscopy and spectroscopy, one can increase the degrees of freedom with the choice of incident and detected wavevector, independent from the active control of the local polarization through an engineered antenna-tip response.
However, to date, most scanning near-field microscopy studies have focused on a surface normal oriented antenna-tip in tip-enhanced near-field microscopy \cite{gerton2004, yano2009, zhang2013chemical} based on the hypothesis of maximum field enhancement in this configuration.

While this conventional tip geometry is useful to selectively detect an \textit{out-of-plane} (tip parallel) polarized response, it reduces the detection sensitivity for \textit{in-plane} (tip perpendicular) polarization.
Artificial tip engineering for enhanced \textit{in-plane} sensitivity limits spatial resolution, universal applicability, and/or requires a complex tip fabrication process \cite{lee2007, olmon2010, burresi2009, kihm2011}.
These limitations in measuring the \textit{in-plane} optical response restrict the range of optical techniques and sample systems. 
Specifically, predominantly two-dimensional (2D) quantum systems, such as graphene \cite{gerber2014, fei2012, chen2012}, transition metal dichalcogenides (TMDs) \cite{park2016tmd, bao2015, li2015}, epitaxial thin films \cite{damodaran2017}, important classes of transition metal oxides of layered materials \cite{kalantar2016}, all with dominant \textit{in-plane} excitations are difficult to probe.
Therefore, to broadly extend the range of near-field microscopy application to the characterization of nano-photonic structures and metasurfaces \cite{kildishev2013}, or optical nano-crystallography and -imaging of anisotropic samples \cite{berweger2009, muller2016}, a new approach with extended antenna-tip vector-field control is desirable.

Here, we demonstrate a generalizable approach to control the excitation and detection polarizability for both \textit{in-plane} and \textit{out-of-plane} vector-fields in nano-imaging, with enhanced sensitivity, and without loss of spatial resolution. 
We break the axial symmetry of a conventional Au wire tip by varying its tilt angle with respect to the sample surface.
By variation of tilt angle, we control the ratio of \textit{in-plane} and \textit{out-of-plane} polarization. 
This oblique angle of the tip axis creates a spatial confinement for free electron oscillation and gives rise to significantly increased field enhancement in both polarization directions resulting from localized surface plasmon resonance (LSPR) effects \cite{talley2005, sanders2016}.

Second-harmonic generation (SHG) microscopy provides structural insight into materials through the nonlinear optical response, such as crystal symmetry, orientation, defect states, stacking angle, and the number of layers \cite{kumar2013, seyler2015, hsu2014}.
We take advantage of the near-field polarization control in both excitation and detection field in symmetry selective tip-enhanced SHG imaging, as an example, applied to different quantum materials.
To quantify the enhanced sensitivity of the tilted tip and to image nano-crystallographic properties, we perform grain boundary (GB) mapping of monolayer MoS$_2$, as a model system of layered 2D materials. 
This is achieved by the reduced nonlinear optical susceptibility and modified selection rule at the grain boundaries.
In addition, on single crystal YMnO$_3$, by mapping both \textit{in-plane} and \textit{out-of-plane} nonlinear optical susceptibility $\chi$$^{(2)}_{ijk}$ components \cite{fiebig2002, neacsu2009}, we obtain ferroelectric domain nano-imaging facilitated by the local phase-sensitive detection \cite{neacsu2009}, with enhanced SHG sensitivity.
These experimental results demonstrate a substantial gain in image content from a simple yet effective modification to the conventional tip-enhanced imaging approach.
The approach is expected to greatly enhance the sensitivity of optial nano-spectroscopy in any linear and nonlinear optical modality and then extends the application space of optical nano-imaging to a wider range of materials.
\\
\begin{figure*}
	\includegraphics[width = 8.5 cm]{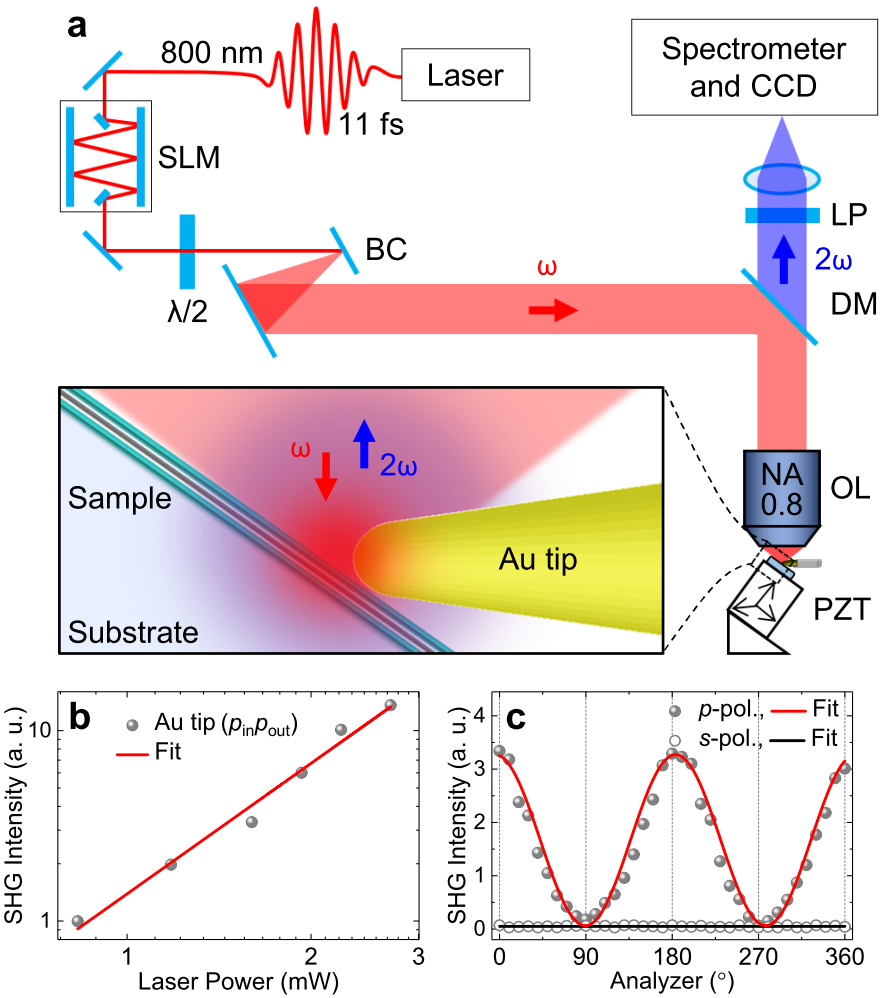}
	\caption{
(a) Schematic of tip-enhanced SHG nano-spectroscopy and -imaging, SLM: spatial light modulator, $\lambda$/2: half wave plate, BC: beam collimator, DM: dichroic mirror, OL: objective lens, LP: linear polarizer (analyzer). 
(b) Log-plot of the power dependence of near-field SHG intensity of Au tip in \textit{p}$_{in}$\textit{p}$_{out}$ configuration. (c) Polarization dependence of SHG intensity of Au tip for \textit{p}- (red) and \textit{s}- (black) polarized excitation.
}
	\label{fig:setup}
\end{figure*}

\noindent
{\bf Experiment}

\noindent
The experiment is based on tip-enhanced spectroscopy \cite{park2016tmd}, with side illumination of the electrochemically etched Au tip manipulated in a shear-force AFM as shown schematically in Fig. 1a.
The sample surface can be tilted by variable angle with respect to the tip axis from 0$^{\circ}$ to 90$^{\circ}$.
Excitation light provided from a Ti:sapphire oscillator (FemtoSource Synergy, Femtolasers Inc., with $\tau$ $\sim$11 fs pulse duration, center wavelength of 800 nm, 78 MHz repetition rate, and $<$ 2 mW power) is focused onto the tip-sample junction using an objective lens (NA = 0.8, W.D. = 4 mm), with polarization and dispersion control.
The backscattered SHG signal is polarization selected and detected using a spectrometer (SpectraPro 500i, Princeton Instruments, f = 500 mm) with a charge-coupled device (CCD) camera (ProEM+: 1600 eXcelon3, Princeton Instruments).

In excitation and detection, we define \textit{p} and \textit{s} polarization as light polarized parallel and perpendicular with respect to the plane formed by \textit{k}-vector and tip axis.
In \textit{p}$_{in}$\textit{p}$_{out}$ (\textit{p} polarized excitation and \textit{p} polarized detection) configuration, the broken axial symmetry gives rise to a tip-SHG response with expected power (Fig. 1b) and polarization (Fig. 1c) dependence of the SHG response.
In SHG nano-imaging, the intrinsic tip-SHG response can be discriminated from the tip-sample coupled response through polarization and tip-sample distance dependent measurements.

Note that a tilted tip geometry has been used in several cases of top-illumination tip-enhanced Raman spectroscopy (TERS) \cite{stadler2010, chan2011}.
However, only for the purpose of ease of tip-illumination without any vector-field control.     
\\
 
\begin{figure*}
	\includegraphics[width = 16.5 cm]{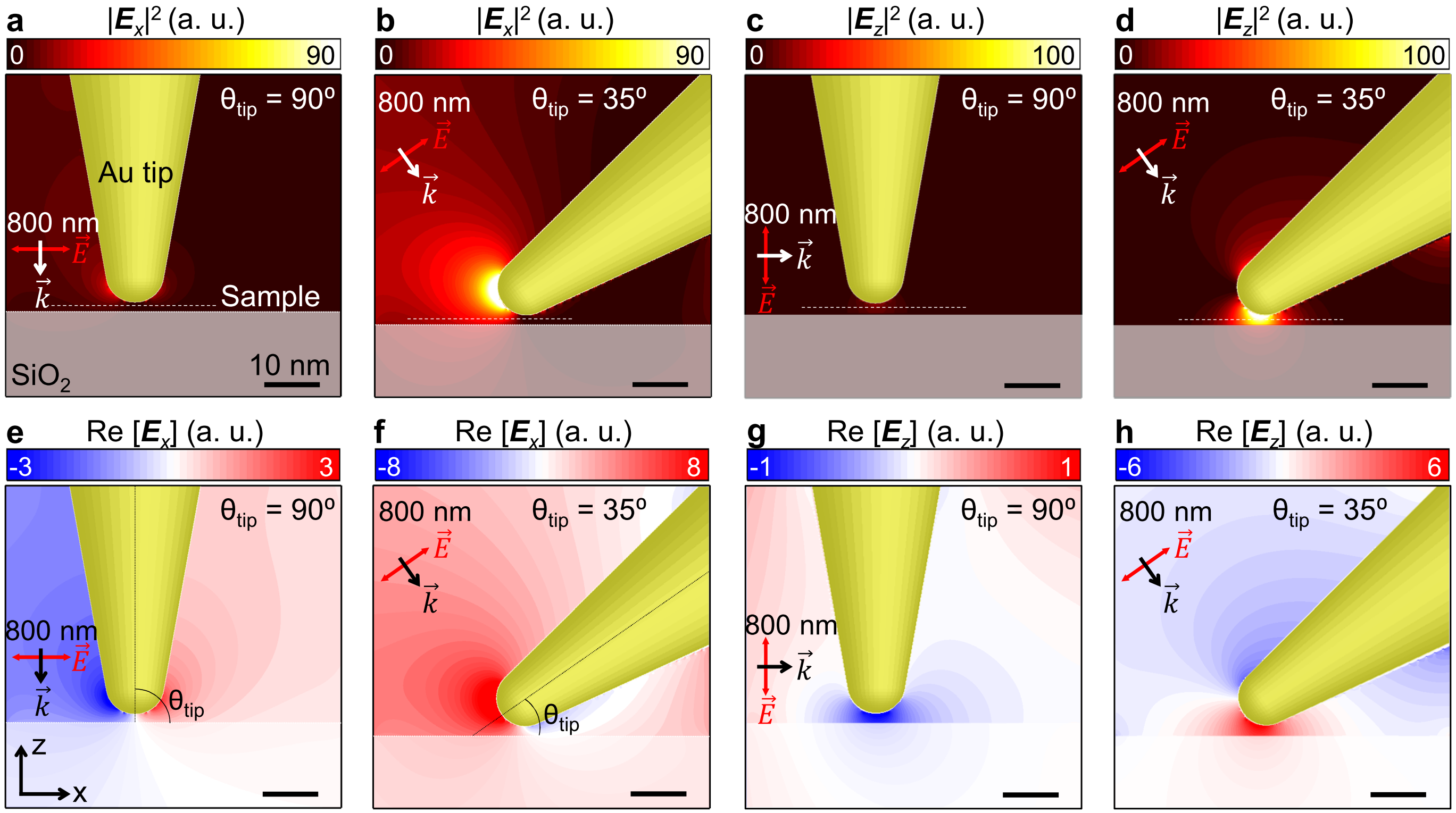}
	\caption{
Simulated optical field distributions for surface normal tip orientation ($\theta$$\rm{_{tip}}$ = 90$^{\circ}$, \textit{in-plane}: $|$\textbf{\textit{E}}$_x$$|$$^2$ (a), Re[\textbf{\textit{E}}$_x$] (e), \textit{out-of-plane}: $|$\textbf{\textit{E}}$_z$$|$$^2$ (c), Re[\textbf{\textit{E}}$_z$] (g)) and for optimal tilted tip orientation ($\theta$$\rm{_{tip}}$ = 35$^{\circ}$, \textit{in-plane}: $|$\textbf{\textit{E}}$_x$$|$$^2$ (b), Re[\textbf{\textit{E}}$_x$] (f), \textit{out-of-plane}: $|$\textbf{\textit{E}}$_z$$|$$^2$ (d), Re[\textbf{\textit{E}}$_z$] (h)). 
}
	\label{fig:tepl}
\end{figure*}

\noindent
{\bf Vector-field control with plasmonic antenna tip}

\noindent 
To characterize the local optical field enhancement with respect to the tilt angle of the Au tip, we calculate the expected optical field distribution using finite-difference time-domain (FDTD) simulations (Lumerical Solutions, Inc.) for our experimental conditions.
Fig. 2a and e show the \textit{in-plane} optical field maps ($|\textbf{\textit{E}}_x|^2$ and Re[\textbf{\textit{E}}$_x$]) for surface normal tip orientation ($\theta$$\rm{_{tip}}$ = 90$^{\circ}$) with an SiO$_2$ substrate, with excitation (800 nm) polarization perpendicular with respect to the tip axis.
A weak \textbf{\textit{E}}$_x$ field confinement at the apex is seen resulting from the transverse local antenna mode \cite{talley2005, sanders2016}. 
\begin{figure*}
	\includegraphics[width = 8.5 cm]{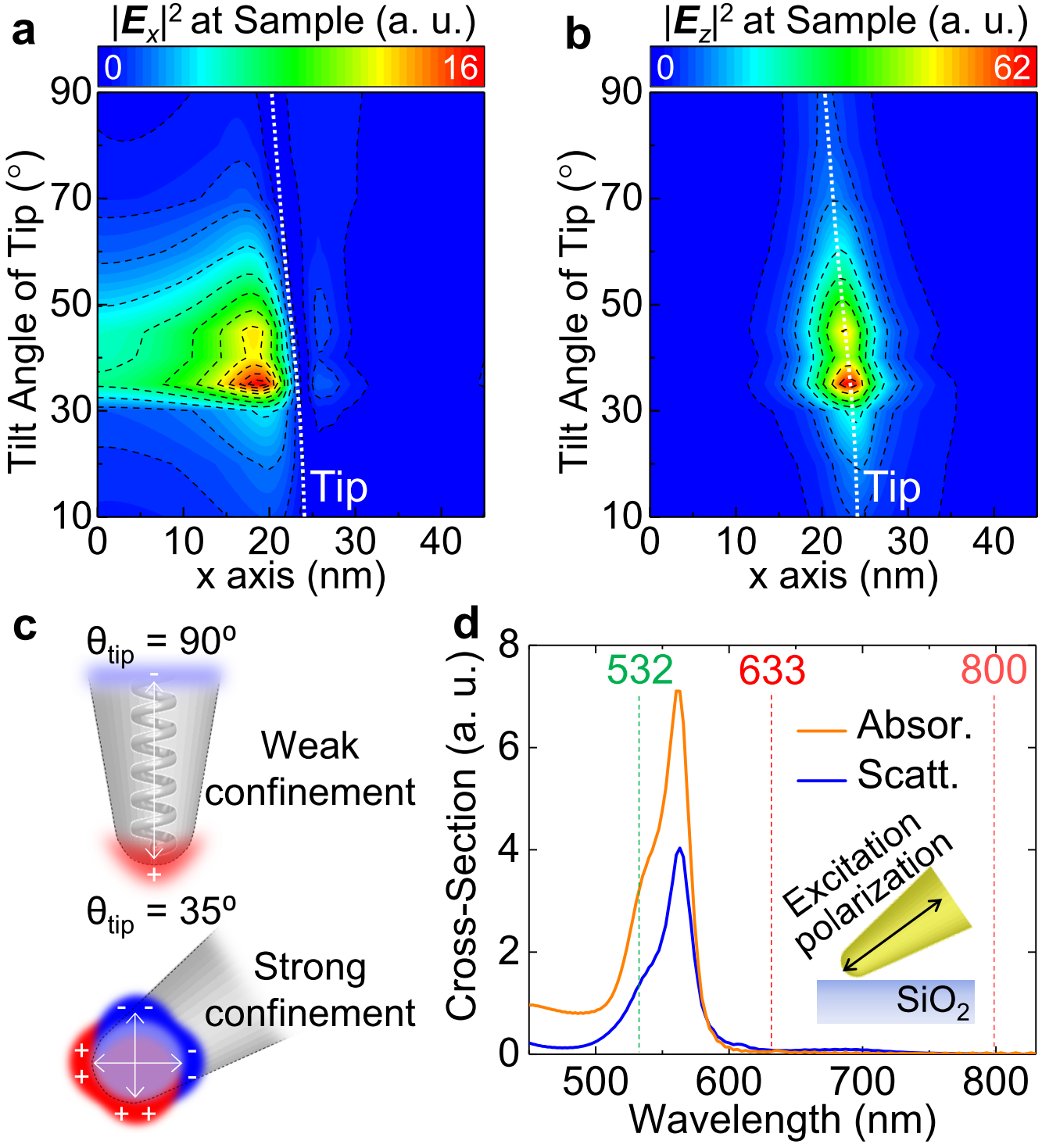}
	\caption{
Simulated \textit{in-plane} $|\textbf{\textit{E}}_x|^2$ (a) and \textit{out-of-plane} $|\textbf{\textit{E}}_z|^2$ (b) optical field intensity profiles at the sample plane with respect to the tilt angle of the tip ($\theta$$\rm{_{tip}}$).
(c) Schematic showing spatial confinement of free electrons oscillation for conventional vertically oriented tip (top, $\theta$$\rm{_{tip}}$ = 90$^{\circ}$) and tilted tip (bottom, $\theta$$\rm{_{tip}}$ = 35$^{\circ}$).
(d) FDTD simulations of the absorption and scattering cross-sections for the tilted Au tip ($\theta$$\rm{_{tip}}$ = 35$^{\circ}$) with SiO$_2$ substrate. 
}
	\label{fig:tepl2}
\end{figure*}
To achieve an efficient local plasmon antenna effect, we model the tilted tip with excitation polarization parallel with respect to the tip axis. 
Fig. 2b and f show calculated $|\textbf{\textit{E}}_x|^2$ and Re[\textbf{\textit{E}}$_x$] distributions for the 35$^{\circ}$ tilted tip orientation ($\theta$$\rm{_{tip}}$ = 35$^{\circ}$), exhibiting $\sim$6 times stronger \textit{in-plane} optical field intensity enhancement compared to sample surface normal orientation (Fig. 2a).
Notably, for this tilt angle, also the \textit{out-of-plane} vector-field is significantly enhanced as seen in Fig. 2c-d and g-h.

To characterize a systematic change of vector-field enhancement, we calculate the \textit{in-plane} and \textit{out-of-plane} optical field intensity with respect to tilt angle.
Fig. 3a and b show simulated vector-field intensity profiles for $|\textbf{\textit{E}}_x|^2$ and $|\textbf{\textit{E}}_z|^2$ at the sample plane (the distance between tip and sample is set to 0.5 nm, see Fig. S1-S9 for full data set of the tilt angle dependent $|\textbf{\textit{E}}_x|^2$ and $|\textbf{\textit{E}}_z|^2$).  
For the small (10$^{\circ}$ $\leq$ $\theta$$\rm{_{tip}}$ $\leq$ 30$^{\circ}$) and large (60$^{\circ}$ $\leq$ $\theta$$\rm{_{tip}}$ $\leq$ 90$^{\circ}$) tilt angles, the field confinement is not significantly enhanced compared to conventional tip orientation ($\theta$$\rm{_{tip}}$ = 90$^{\circ}$) due to the overdamped resonance of the electrons oscillation in a semi-infinite tip structure.
In this case, the Au tip cannot sustain antenna-like \textit{in-plane} and \textit{out-of-plane} surface plasmon polaritons (SPPs) \cite{sanders2016}.
On the other hand, the field confinement is significantly enhanced for the tilt angles between 30$^{\circ}$ and 60$^{\circ}$ because geometrically confined free electrons give rise to an appreciable LSPR effect, as illustrated in Fig. 3c. 
Note that the Au tip with the SiO$_2$ substrate provides for larger vector-field enhancement than the free standing tip because the SiO$_2$ substrate gives rise to an induced dipole coupling between the tip and sample (see Fig. S1-S9, discussing the substrate effect) \cite{notingher2005}.

\begin{table} [t!]
\centering
\begin{tabular}{|  c  |  c  |  c  |  c  |  c  |}
\hline 
\multirow{2}{*}{} & \multicolumn{2}{c|}{$|\textbf{\textit{E}}_x/\textbf{\textit{E}}_0|^2$ (\textit{in-plane})} & \multicolumn{2}{c|}{$|\textbf{\textit{E}}_z/\textbf{\textit{E}}_0|^2$ (\textit{out-of-plane})} \\ \cline{2-5} 
                  & $\theta$$\rm{_{tip}}$ = 90$^{\circ}$ & $\theta$$\rm{_{tip}}$ = 35$^{\circ}$ & $\theta$$\rm{_{tip}}$ = 90$^{\circ}$ & $\theta$$\rm{_{tip}}$ = 35$^{\circ}$  \\ \hline
 $\lambda$$\rm{_{exc}}$ = 532 nm & 28 & 290 & 40 & 630  \\ \hline
 $\lambda$$\rm{_{exc}}$ = 633 nm & 23 & 180 & 42 & 250  \\ \hline
 $\lambda$$\rm{_{exc}}$ = 800 nm & 16 & 90 & 25 & 100  \\ \hline 
\end{tabular}
\caption[Optical field intensity enhancement of tilted tip]{Comparison of \textit{in-plane} and \textit{out-of-plane} optical field intensity enhancement for conventional ($\theta$$\rm{_{tip}}$ = 90$^{\circ}$) and tilted ($\theta$$\rm{_{tip}}$ = 35$^{\circ}$) tip for selected excitation wavelengths, {color{red}exhibiting larger field enhancement of the tilted tip at the resonance excitation for both $|\textbf{\textit{E}}_x|^2$ and $|\textbf{\textit{E}}_z|^2$.}}
\label{tab:tilt}
\end{table}

Fig. 3d shows calculated absorption and scattering cross-section spectra for the tilted Au tip ($\theta$$\rm{_{tip}}$ = 35$^{\circ}$) near the SiO$_2$ substrate.
The LSPR of the tilted tip is at $\sim$550 nm near the interband transition of gold (2.4 eV), with modified spectral shape and linewidth due to the elongated structure, and correspondingly modfied radiative damping \cite{grigorchuk2012}.
Table~\ref{tab:tilt} shows comparisons of the resulting \textit{in-plane} and \textit{out-of-plane} vector-field intensity enhancement for conventional tip ($\theta$$\rm{_{tip}}$ = 90$^{\circ}$) and tilted tip ($\theta$$\rm{_{tip}}$ = 35$^{\circ}$) for selected on and off resonance excitation wavelengths.
As can be seen, the tilted tip results in a much larger optical field enhancement for both $|\textbf{\textit{E}}_x/\textbf{\textit{E}}_0|^2$ ($\textbf{\textit{I}}$$^{35^\circ}_{\omega}$/$\textbf{\textit{I}}$$^{90^\circ}_{\omega}$ $\sim$ 6-10) and $|\textbf{\textit{E}}_z/\textbf{\textit{E}}_0|^2$ ($\textbf{\textit{I}}$$^{35^\circ}_{\omega}$/$\textbf{\textit{I}}$$^{90^\circ}_{\omega}$ $\sim$ 4-16) for all wavelengths, with the largest effect on resonance.
Based on these results, we understand that the tilted tip induces a strongly localized plasmon resonance to both \textit{in-plane} and \textit{out-of-plane} directions by creating a spatial confinement for free electrons oscillation in contrast to a reduced resonance effect for surface normal tip orientation.
\\

\begin{figure*}
	\includegraphics[width = 16.5 cm]{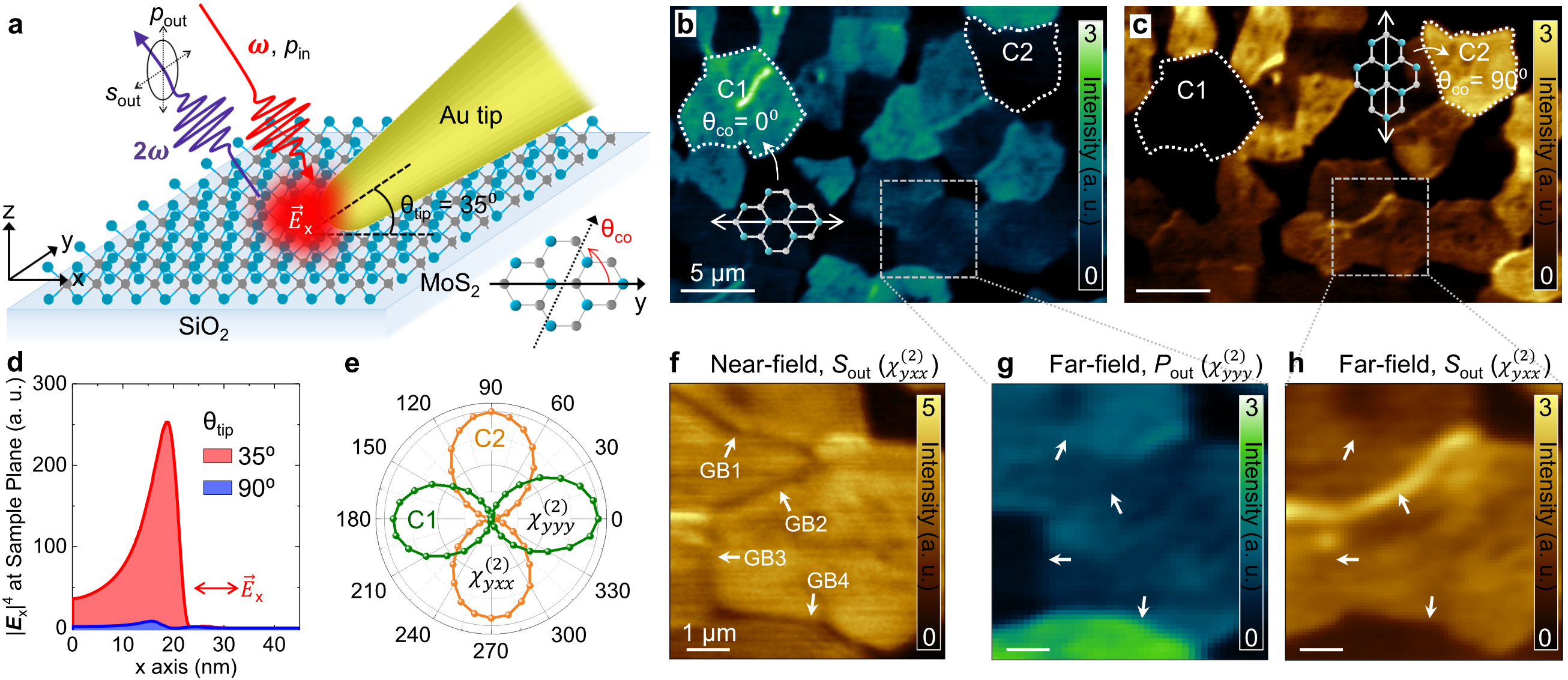}
	\caption{
(a) Schematic of tip-enhanced SHG nano-crystallography imaging for monolayer MoS$_2$ films on a SiO$_2$ substrate. 
Far-field SHG images measured in \textit{p}$_{in}$\textit{p}$_{out}$ (b) and \textit{p}$_{in}$\textit{s}$_{out}$ (c) configurations. 
(d) Simulated $|$\textbf{\textit{E}}$_x$$|$$^4$ profile at sample plane for tilted ($\theta$$\rm{_{tip}}$ = 35$^{\circ}$) and conventional ($\theta$$\rm{_{tip}}$ = 90$^{\circ}$) tip orientations. 
(e) Far-field SHG polarization dependence of crystals C1 and C2 in (b) and (c).
(f) Tip-enhanced SHG nano-crystallographic image of the same area measured in \textit{p}$_{in}$\textit{s}$_{out}$ configuration. 
(g)-(h) Magnified far-field SHG images of small area in (b)-(c). 
}
	\label{fig:spec}
\end{figure*}

\noindent
{\bf Nonlinear optical nano-crystallography and nano-imaging}

\noindent
As illustrated in Fig. 4a, we perform tip-enhanced nano-SHG imaging with the 35$^{\circ}$ tilted antenna-tip for single-layer MoS$_2$ films grown on a SiO$_2$/Si substrate, as a model system of \textit{$\bar{6}$m2} point group possessing pure \textit{in-plane} $\chi$$^{(2)}_{ijk}$ tensor elements \cite{malard2013, li2013, kumar2013}.

For comparison, we first perform conventional far-field imaging to determine crystal orientation angle ($\theta$$\rm{_{co}}$) and grain boundary of the TMD crystals \cite{yin2014, cheng2015}.
Fig. 4b and c show far-field SHG images with polarization selections of \textit{p}$_{in}$\textit{p}$_{out}$ and \textit{p}$_{in}$\textit{s}$_{out}$, respectively.
From the nonvanishing $\chi$$^{(2)}_{ijk}$ tensor elements and excitation condition, the induced second-order polarization for crystals with $\theta$$\rm{_{co}}$ = 0$^{\circ}$ (C1 of Fig. 4b-c) is given by $\textbf{\textit{P}}$$_y$(2$\omega$) = 2$\varepsilon_0$$\chi$$^{(2)}_{yyy}$$\textbf{\textit{E}}$$_y$($\omega$)$^2$,
where $\textbf{\textit{E}}$$_{i=x, y, z}$($\omega$) are the electric field components at the laser frequency (see Supporting Information for detailed matrix representations and calculations).
Therefore, the SHG signal of crystals with $\theta$$\rm{_{co}}$ = 0$^{\circ}$ is polarized parallel to the excitation polarization ($\omega$), and these crystals are clearly observed in \textit{p}$_{in}$\textit{p}$_{out}$ configuration, as shown in Fig. 4b.
In contrast, crystals with $\theta$$\rm{_{co}}$ = 90$^{\circ}$ (C2 of Fig. 4b-c) are seen most clearly in \textit{p}$_{in}$\textit{s}$_{out}$ configuration (Fig. 4c) since the induced SHG polarization is given by $\textbf{\textit{P}}$$_y$(2$\omega$) = -2$\varepsilon_0$$\chi$$^{(2)}_{yxx}$$\textbf{\textit{E}}$$_x$($\omega$)$^2$.
This polarization dependence on crystallographic orientation is also confirmed in far-field SHG anisotropy measured with rotating analyzer (Fig. 4e).

Only $\textbf{\textit{E}}$$_{x}$($\omega$) contributes to the SHG signal in \textit{p}$_{in}$\textit{s}$_{out}$ configuration.
Therefore, with $\textbf{\textit{I}}$$_{2\omega}$ $\propto$ $|$\textbf{\textit{P}}(2$\omega$)$|$$^2$ $\propto$ $|$\textbf{\textit{E}}($\omega$)$|$$^4$,
we can calculate the enhanced SHG intensity using the 35$^{\circ}$ tilted tip ($\textbf{\textit{I}}$$^{35^\circ}_{2\omega, \textup{MoS$_2$}}$) compared to the conventional surface normal oriented tip ($\textbf{\textit{I}}$$^{90^\circ}_{2\omega, \textup{MoS$_2$}}$) from the FDTD simulation.
As shown in Fig. 4d, the spatially integrated $|$$\textbf{\textit{E}}$$_x$($\omega$)$|$$^4$ for the 35$^{\circ}$ tilted tip at the sample plane is ${\sim}$28 times larger than that of the surface normal oriented tip ($\theta$$\rm{_{tip}}$ = 90$^{\circ}$), i.e., $\textbf{\textit{I}}$$^{35^\circ}_{2\omega, \textup{MoS$_2$}}$/$\textbf{\textit{I}}$$^{90^\circ}_{2\omega, \textup{MoS$_2$}}$ $\sim$ 28.

Fig. 4f shows a measured tip-enhanced nano-SHG image in \textit{p}$_{in}$\textit{s}$_{out}$ configuration for a small area selected within the far-field image.
Far-field images of \textit{p}$_{out}$ and \textit{s}$_{out}$ detection are magnified in Fig. 4g and h for comparison.
As demonstrated previously \cite{yin2014, cheng2015}, some GBs are visualized in the far-field SHG images due to the constructive (or destructive) interference between SHG signals of adjacent crystals.
However, this interference contrast at GBs is observed only for specific crystal orientation or polarization condition \cite{cheng2015}.

In contrast, in the tip-enhanced SHG image (Fig. 4f), a full GB map is obtained with pronounced SHG contrast.
For example, while GB2 is observed in both far- and near-field images, the additional GBs (GB1, GB3, and GB4) are only seen in the near-field image.
In contrast to the far-field response, a full GB map can be obtained regardless of crystal orientation and interference.

\begin{figure}
	\includegraphics[width = 16 cm]{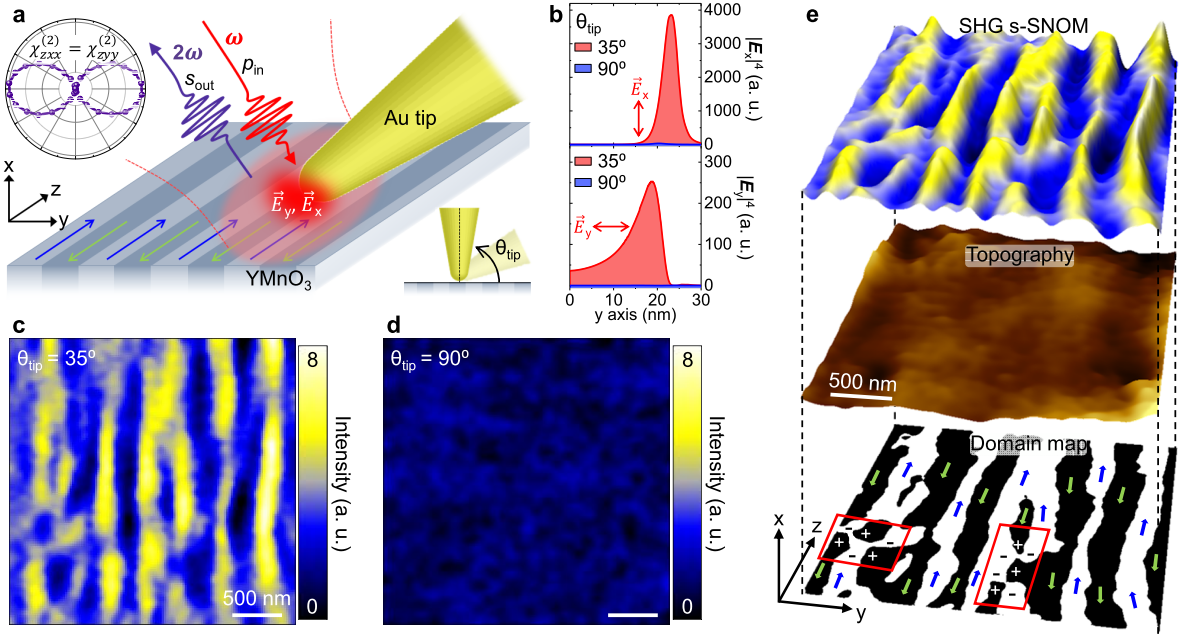}
	\caption{
(a) Schematic of tip-enhanced SHG nano-crystallography imaging for single-crystalline \textit{x}-cut YMnO$_3$. 
(b) Simulated $|$\textbf{\textit{E}}$_x$$|$$^4$ (\textit{out-of-plane}) and $|$\textbf{\textit{E}}$_y$$|$$^4$ (\textit{in-plane}) profiles at sample plane for tilted ($\theta$$\rm{_{tip}}$ = 35$^{\circ}$) and conventional ($\theta$$\rm{_{tip}}$ = 90$^{\circ}$) tip orientations. 
Tip-enhanced SHG nano-crystallographic image measured by tilted (c, $\theta$$\rm{_{tip}}$ = 35$^{\circ}$) and conventional (d, $\theta$$\rm{_{tip}}$ = 90$^{\circ}$) tip. 
(e) Three dimensional representation of SHG nano-crystallographic image and topography, together with the corresponding ferroelectric domain map.
}
	\label{fig:edge}
\end{figure}

In order to assess the full benefit of increased both \textit{in-} and \textit{out-of-plane} field confinement (Fig. 2d), we then perform tip-enhanced nano-SHG imaging on single-crystalline \textit{x}-cut YMnO$_3$, as a model system of \textit{6mm} point group with both \textit{in-plane} and \textit{out-of-plane} nonlinear optical susceptibility \cite{fiebig2002, neacsu2009}.
We first deduce the microscopic sample orientation from far-field SHG anisotropy measurement, as shown in Fig. 5a.
Based on this information, we probe the ferroelectric $\chi$$^{(2)}_{zxx}$ = $\chi$$^{(2)}_{zyy}$ tensor elements in \textit{p}$_{in}$\textit{s}$_{out}$ tip-enhanced near-field microscopy configuration.
The corresponding SHG polarization is then given by 
$\textbf{\textit{P}}$$_z$(2$\omega$) = 2$\varepsilon_0$$\chi$$^{(2)}_{zxx}$($\textbf{\textit{E}}$$_x$($\omega$)$^2$+$\textbf{\textit{E}}$$_y$($\omega$)$^2$) (see Supporting Information for detailed matrix representations and calculations).
The measured intensity 
$\textbf{\textit{I}}$$_{2\omega}$
is proportional to 
$|$$\textbf{\textit{E}}$$_x$($\omega$)$^2$+$\textbf{\textit{E}}$$_y$($\omega$)$^2$$|$$^2$ = $|$$\textbf{\textit{E}}$$_x$($\omega$)$|$$^4$ + $|$$\textbf{\textit{E}}$$_y$($\omega$)$|$$^4$ + 2$|$$\textbf{\textit{E}}$$_x$($\omega$)$|$$^2$$|$$\textbf{\textit{E}}$$_y$($\omega$)$|$$^2$.
From the spatially integrated 
$|$$\textbf{\textit{E}}$$_x$($\omega$)$|$$^4$, $|$$\textbf{\textit{E}}$$_y$($\omega$)$|$$^4$, $|$$\textbf{\textit{E}}$$_x$($\omega$)$|$$^2$, and $|$$\textbf{\textit{E}}$$_y$($\omega$)$|$$^2$
values at the sample plane for the 35$^{\circ}$ tilted and surface normal oriented tips (Fig. 5b), we can estimate the tip-enhanced SHG intensity ratio of $\textbf{\textit{I}}$$^{35^\circ}_{2\omega, \textup{YMnO$_3$}}$/$\textbf{\textit{I}}$$^{90^\circ}_{2\omega, \textup{YMnO$_3$}}$ $\sim$ 27.

Fig. 5c and d show the resulting tip-enhanced SHG images under tilted tip ($\theta$$\rm{_{tip}}$ = 35$^{\circ}$) and surface normal tip ($\theta$$\rm{_{tip}}$ = 90$^{\circ}$) configuration.
We observe a high contrast image of the domains only with the tilted tip.
The details of contrast are due to the interference between the tip-enhanced SHG from a single domain and residual far-field SHG from multiple domains giving rise to a local phase-sensitive signal \cite{neacsu2009}.
From that image, we can obtain the corresponding ferroelectric domain map exhibiting an alternating ferroelectric polarization pattern as expected for this crystallographic orientation (Fig. 5e).
In addition, we observe the 3-fold symmetric vortices of the domains (red boxes) as expected for hexagonal manganites \cite{chae2012, jungk2010} which provides information for the understanding of topological behaviors of ferroics.

In summary, a conventional surface normal oriented tip geometry in tip-enhanced near-field microscopy gives limited polarization control in both the intrinsic far-field excitation and the extrinsic near-field nano-optical response.
Furthermore, for surface normal tip orientation, the antenna mode driven into a semi-infinite tip structure results in reduced field enhancement due to overdamping, which gives rise to reduced efficiency for both \textit{in-plane} and \textit{out-of-plane} nano-optical response. 
Our work presents a simple but powerful solution to control the vector-field of a nano-optical antenna-tip.
We show that the optical field confinement can be systematically controlled by tuning the tip orientation angle with respect to the sample surface, to enhance the \textit{in-plane} optical field (\textbf{\textit{E}}$_x$) confinement for investigation of 2D materials. 
Surprisingly, rather than an associated decrease in \textit{out-of-plane} sensitivity with increasing tilt angle, the \textit{out-of-plane} optical field (\textbf{\textit{E}}$_z$) is also enhanced with an even larger enhancement factor than \textbf{\textit{E}}$_x$.
We find that at an optimized angle near 35$^{\circ}$ with details depending on tip material, sample, and excitation wavelength, the broken axial symmetry provides for a more sensitive nano-probe beyond conventional near-field microscopy tip for all optical modalities and any sample.
The vector-field controllability of plasmonic antenna-tip not only allows probing selective polarization components of samples by simply changing its tilting angle but also this strongly confined vector-field gives access to anomalous nanoscale light-matter interactions such as exciton-plasmon coupling \cite{park2017dark}, electron-phonon coupling \cite{jin2016}, and strong coupling \cite{chikkaraddy2016} in a range of photoactive molecules and quantum materials.
\\

\bibliography{TiltSHG} 

\begin{thebibliography}{10}
\expandafter\ifx\csname url\endcsname\relax
  \def\url#1{\texttt{#1}}\fi
\expandafter\ifx\csname urlprefix\endcsname\relax\def\urlprefix{URL }\fi
\providecommand{\bibinfo}[2]{#2}
\providecommand{\eprint}[2][]{\url{#2}}

\bibitem{anastassakis1997}
\bibinfo{author}{Anastassakis, E.}
\newblock \bibinfo{title}{Selection rules of raman scattering by optical
  phonons in strained cubic crystals}.
\newblock \emph{\bibinfo{journal}{J. Appl. Phys.}}
  \textbf{\bibinfo{volume}{82}}, \bibinfo{pages}{1582--1591}
  (\bibinfo{year}{1997}).

\bibitem{fiebig2000}
\bibinfo{author}{Fiebig, M.} \emph{et~al.}
\newblock \bibinfo{title}{Determination of the magnetic symmetry of hexagonal
  manganites by second harmonic generation}.
\newblock \emph{\bibinfo{journal}{Phys. Rev. Lett.}}
  \textbf{\bibinfo{volume}{84}}, \bibinfo{pages}{5620} (\bibinfo{year}{2000}).

\bibitem{najafov2010}
\bibinfo{author}{Najafov, H.} \emph{et~al.}
\newblock \bibinfo{title}{Observation of long-range exciton diffusion in highly
  ordered organic semiconductors}.
\newblock \emph{\bibinfo{journal}{Nat. Mater.}} \textbf{\bibinfo{volume}{9}},
  \bibinfo{pages}{938} (\bibinfo{year}{2010}).

\bibitem{yin2014}
\bibinfo{author}{Yin, X.} \emph{et~al.}
\newblock \bibinfo{title}{Edge nonlinear optics on a {M}o{S}$_2$ atomic
  monolayer}.
\newblock \emph{\bibinfo{journal}{Science}} \textbf{\bibinfo{volume}{344}},
  \bibinfo{pages}{488--490} (\bibinfo{year}{2014}).

\bibitem{gerton2004}
\bibinfo{author}{Gerton, J.~M.}, \bibinfo{author}{Wade, L.~A.},
  \bibinfo{author}{Lessard, G.~A.}, \bibinfo{author}{Ma, Z.} \&
  \bibinfo{author}{Quake, S.~R.}
\newblock \bibinfo{title}{Tip-enhanced fluorescence microscopy at 10 nanometer
  resolution}.
\newblock \emph{\bibinfo{journal}{Phys. Rev. Lett.}}
  \textbf{\bibinfo{volume}{93}}, \bibinfo{pages}{180801}
  (\bibinfo{year}{2004}).

\bibitem{yano2009}
\bibinfo{author}{Yano, T.-a.}, \bibinfo{author}{Verma, P.},
  \bibinfo{author}{Saito, Y.}, \bibinfo{author}{Ichimura, T.} \&
  \bibinfo{author}{Kawata, S.}
\newblock \bibinfo{title}{Pressure-assisted tip-enhanced raman imaging at a
  resolution of a few nanometres}.
\newblock \emph{\bibinfo{journal}{Nat. Photon.}} \textbf{\bibinfo{volume}{3}},
  \bibinfo{pages}{473--477} (\bibinfo{year}{2009}).

\bibitem{zhang2013chemical}
\bibinfo{author}{Zhang, R.} \emph{et~al.}
\newblock \bibinfo{title}{Chemical mapping of a single molecule by
  plasmon-enhanced raman scattering}.
\newblock \emph{\bibinfo{journal}{Nature}} \textbf{\bibinfo{volume}{498}},
  \bibinfo{pages}{82--86} (\bibinfo{year}{2013}).

\bibitem{lee2007}
\bibinfo{author}{Lee, K.} \emph{et~al.}
\newblock \bibinfo{title}{Vector field microscopic imaging of light}.
\newblock \emph{\bibinfo{journal}{Nat. Photon.}} \textbf{\bibinfo{volume}{1}},
  \bibinfo{pages}{53--56} (\bibinfo{year}{2007}).

\bibitem{olmon2010}
\bibinfo{author}{Olmon, R.~L.} \emph{et~al.}
\newblock \bibinfo{title}{Determination of electric-field, magnetic-field, and
  electric-current distributions of infrared optical antennas: a near-field
  optical vector network analyzer}.
\newblock \emph{\bibinfo{journal}{Phys. Rev. Lett.}}
  \textbf{\bibinfo{volume}{105}}, \bibinfo{pages}{167403}
  (\bibinfo{year}{2010}).

\bibitem{burresi2009}
\bibinfo{author}{Burresi, M.} \emph{et~al.}
\newblock \bibinfo{title}{Probing the magnetic field of light at optical
  frequencies}.
\newblock \emph{\bibinfo{journal}{Science}} \textbf{\bibinfo{volume}{326}},
  \bibinfo{pages}{550--553} (\bibinfo{year}{2009}).

\bibitem{kihm2011}
\bibinfo{author}{Kihm, H.} \emph{et~al.}
\newblock \bibinfo{title}{Bethe-hole polarization analyser for the magnetic
  vector of light}.
\newblock \emph{\bibinfo{journal}{Nat. Commun.}} \textbf{\bibinfo{volume}{2}},
  \bibinfo{pages}{451} (\bibinfo{year}{2011}).

\bibitem{gerber2014}
\bibinfo{author}{Gerber, J.~A.}, \bibinfo{author}{Berweger, S.},
  \bibinfo{author}{O’Callahan, B.~T.} \& \bibinfo{author}{Raschke, M.~B.}
\newblock \bibinfo{title}{Phase-resolved surface plasmon interferometry of
  graphene}.
\newblock \emph{\bibinfo{journal}{Phys. Rev. Lett.}}
  \textbf{\bibinfo{volume}{113}}, \bibinfo{pages}{055502}
  (\bibinfo{year}{2014}).

\bibitem{fei2012}
\bibinfo{author}{Fei, Z.} \emph{et~al.}
\newblock \bibinfo{title}{Gate-tuning of graphene plasmons revealed by infrared
  nano-imaging}.
\newblock \emph{\bibinfo{journal}{Nature}} \textbf{\bibinfo{volume}{487}},
  \bibinfo{pages}{82} (\bibinfo{year}{2012}).

\bibitem{chen2012}
\bibinfo{author}{Chen, J.} \emph{et~al.}
\newblock \bibinfo{title}{Optical nano-imaging of gate-tunable graphene
  plasmons}.
\newblock \emph{\bibinfo{journal}{Nature}} \textbf{\bibinfo{volume}{487}},
  \bibinfo{pages}{77--81} (\bibinfo{year}{2012}).

\bibitem{park2016tmd}
\bibinfo{author}{Park, K.-D.} \emph{et~al.}
\newblock \bibinfo{title}{Hybrid tip-enhanced nanospectroscopy and nanoimaging
  of monolayer {WS}e$_2$ with local strain control}.
\newblock \emph{\bibinfo{journal}{Nano Lett.}} \textbf{\bibinfo{volume}{16}},
  \bibinfo{pages}{2621--2627} (\bibinfo{year}{2016}).

\bibitem{bao2015}
\bibinfo{author}{Bao, W.} \emph{et~al.}
\newblock \bibinfo{title}{Visualizing nanoscale excitonic relaxation properties
  of disordered edges and grain boundaries in monolayer molybdenum disulfide}.
\newblock \emph{\bibinfo{journal}{Nat. Commun.}} \textbf{\bibinfo{volume}{6}},
  \bibinfo{pages}{8993} (\bibinfo{year}{2015}).

\bibitem{li2015}
\bibinfo{author}{Li, Z.} \emph{et~al.}
\newblock \bibinfo{title}{Active light control of the {M}o{S}$_2$ monolayer
  exciton binding energy}.
\newblock \emph{\bibinfo{journal}{ACS Nano}} \textbf{\bibinfo{volume}{9}},
  \bibinfo{pages}{10158--10164} (\bibinfo{year}{2015}).

\bibitem{damodaran2017}
\bibinfo{author}{Damodaran, A.} \emph{et~al.}
\newblock \bibinfo{title}{Phase coexistence and electric-field control of
  toroidal order in oxide superlattices}.
\newblock \emph{\bibinfo{journal}{Nat. Mater.}}  (\bibinfo{year}{2017}).

\bibitem{kalantar2016}
\bibinfo{author}{Kalantar-zadeh, K.} \emph{et~al.}
\newblock \bibinfo{title}{Two dimensional and layered transition metal oxides}.
\newblock \emph{\bibinfo{journal}{Appl. Mater. Today}}
  \textbf{\bibinfo{volume}{5}}, \bibinfo{pages}{73--89} (\bibinfo{year}{2016}).

\bibitem{kildishev2013}
\bibinfo{author}{Kildishev, A.~V.}, \bibinfo{author}{Boltasseva, A.} \&
  \bibinfo{author}{Shalaev, V.~M.}
\newblock \bibinfo{title}{Planar photonics with metasurfaces}.
\newblock \emph{\bibinfo{journal}{Science}} \textbf{\bibinfo{volume}{339}},
  \bibinfo{pages}{1232009} (\bibinfo{year}{2013}).

\bibitem{berweger2009}
\bibinfo{author}{Berweger, S.} \emph{et~al.}
\newblock \bibinfo{title}{Optical nanocrystallography with tip-enhanced phonon
  raman spectroscopy}.
\newblock \emph{\bibinfo{journal}{Nat. Nanotech.}}
  \textbf{\bibinfo{volume}{4}}, \bibinfo{pages}{496--499}
  (\bibinfo{year}{2009}).

\bibitem{muller2016}
\bibinfo{author}{Muller, E.~A.}, \bibinfo{author}{Pollard, B.},
  \bibinfo{author}{Bechtel, H.~A.}, \bibinfo{author}{van Blerkom, P.} \&
  \bibinfo{author}{Raschke, M.~B.}
\newblock \bibinfo{title}{Infrared vibrational nanocrystallography and
  nanoimaging}.
\newblock \emph{\bibinfo{journal}{Sci. Adv.}} \textbf{\bibinfo{volume}{2}},
  \bibinfo{pages}{e1601006} (\bibinfo{year}{2016}).

\bibitem{talley2005}
\bibinfo{author}{Talley, C.~E.} \emph{et~al.}
\newblock \bibinfo{title}{Surface-enhanced raman scattering from individual
  {A}u nanoparticles and nanoparticle dimer substrates}.
\newblock \emph{\bibinfo{journal}{Nano Lett.}} \textbf{\bibinfo{volume}{5}},
  \bibinfo{pages}{1569--1574} (\bibinfo{year}{2005}).

\bibitem{sanders2016}
\bibinfo{author}{Sanders, A.} \emph{et~al.}
\newblock \bibinfo{title}{Understanding the plasmonics of nanostructured atomic
  force microscopy tips}.
\newblock \emph{\bibinfo{journal}{Appl. Phys. Lett.}}
  \textbf{\bibinfo{volume}{109}}, \bibinfo{pages}{153110}
  (\bibinfo{year}{2016}).

\bibitem{kumar2013}
\bibinfo{author}{Kumar, N.} \emph{et~al.}
\newblock \bibinfo{title}{Second harmonic microscopy of monolayer {M}o{S}$_2$}.
\newblock \emph{\bibinfo{journal}{Phys. Rev. B}} \textbf{\bibinfo{volume}{87}},
  \bibinfo{pages}{161403} (\bibinfo{year}{2013}).

\bibitem{seyler2015}
\bibinfo{author}{Seyler, K.~L.} \emph{et~al.}
\newblock \bibinfo{title}{Electrical control of second-harmonic generation in a
  {WS}e$_2$ monolayer transistor}.
\newblock \emph{\bibinfo{journal}{Nat. Nanotech.}}
  \textbf{\bibinfo{volume}{10}}, \bibinfo{pages}{407--411}
  (\bibinfo{year}{2015}).

\bibitem{hsu2014}
\bibinfo{author}{Hsu, W.-T.} \emph{et~al.}
\newblock \bibinfo{title}{Second harmonic generation from artificially stacked
  transition metal dichalcogenide twisted bilayers}.
\newblock \emph{\bibinfo{journal}{ACS Nano}} \textbf{\bibinfo{volume}{8}},
  \bibinfo{pages}{2951--2958} (\bibinfo{year}{2014}).

\bibitem{fiebig2002}
\bibinfo{author}{Fiebig, M.}, \bibinfo{author}{Fr{\"o}hlich, D.},
  \bibinfo{author}{Lottermoser, T.} \& \bibinfo{author}{Maat, M.}
\newblock \bibinfo{title}{Probing of ferroelectric surface and bulk domains in
  {RM}n{O}$_3$ ({R} = {Y}, {H}o) by second harmonic generation}.
\newblock \emph{\bibinfo{journal}{Phys. Rev. B}} \textbf{\bibinfo{volume}{66}},
  \bibinfo{pages}{144102} (\bibinfo{year}{2002}).

\bibitem{neacsu2009}
\bibinfo{author}{Neacsu, C.~C.}, \bibinfo{author}{van Aken, B.~B.},
  \bibinfo{author}{Fiebig, M.} \& \bibinfo{author}{Raschke, M.~B.}
\newblock \bibinfo{title}{Second-harmonic near-field imaging of ferroelectric
  domain structure of {YM}n{O}$_3$}.
\newblock \emph{\bibinfo{journal}{Phys. Rev. B}} \textbf{\bibinfo{volume}{79}},
  \bibinfo{pages}{100107} (\bibinfo{year}{2009}).

\bibitem{stadler2010}
\bibinfo{author}{Stadler, J.}, \bibinfo{author}{Schmid, T.} \&
  \bibinfo{author}{Zenobi, R.}
\newblock \bibinfo{title}{Nanoscale chemical imaging using top-illumination
  tip-enhanced raman spectroscopy}.
\newblock \emph{\bibinfo{journal}{Nano Lett.}} \textbf{\bibinfo{volume}{10}},
  \bibinfo{pages}{4514--4520} (\bibinfo{year}{2010}).

\bibitem{chan2011}
\bibinfo{author}{Chan, K.~A.} \& \bibinfo{author}{Kazarian, S.~G.}
\newblock \bibinfo{title}{Tip-enhanced raman mapping with top-illumination
  afm}.
\newblock \emph{\bibinfo{journal}{Nanotechnology}}
  \textbf{\bibinfo{volume}{22}}, \bibinfo{pages}{175701}
  (\bibinfo{year}{2011}).

\bibitem{notingher2005}
\bibinfo{author}{Notingher, I.} \& \bibinfo{author}{Elfick, A.}
\newblock \bibinfo{title}{Effect of sample and substrate electric properties on
  the electric field enhancement at the apex of spm nanotips}.
\newblock \emph{\bibinfo{journal}{J. Phys. Chem. B}}
  \textbf{\bibinfo{volume}{109}}, \bibinfo{pages}{15699--15706}
  (\bibinfo{year}{2005}).

\bibitem{grigorchuk2012}
\bibinfo{author}{Grigorchuk, N.~I.}
\newblock \bibinfo{title}{Radiative damping of surface plasmon resonance in
  spheroidal metallic nanoparticle embedded in a dielectric medium}.
\newblock \emph{\bibinfo{journal}{JOSA B}} \textbf{\bibinfo{volume}{29}},
  \bibinfo{pages}{3404--3411} (\bibinfo{year}{2012}).

\bibitem{malard2013}
\bibinfo{author}{Malard, L.~M.}, \bibinfo{author}{Alencar, T.~V.},
  \bibinfo{author}{Barboza, A. P.~M.}, \bibinfo{author}{Mak, K.~F.} \&
  \bibinfo{author}{de~Paula, A.~M.}
\newblock \bibinfo{title}{Observation of intense second harmonic generation
  from {M}o{S}$_2$ atomic crystals}.
\newblock \emph{\bibinfo{journal}{Phys. Rev. B}} \textbf{\bibinfo{volume}{87}},
  \bibinfo{pages}{201401} (\bibinfo{year}{2013}).

\bibitem{li2013}
\bibinfo{author}{Li, Y.} \emph{et~al.}
\newblock \bibinfo{title}{Probing symmetry properties of few-layer {M}o{S}$_2$
  and h-{BN} by optical second-harmonic generation}.
\newblock \emph{\bibinfo{journal}{Nano Lett.}} \textbf{\bibinfo{volume}{13}},
  \bibinfo{pages}{3329--3333} (\bibinfo{year}{2013}).

\bibitem{cheng2015}
\bibinfo{author}{Cheng, J.} \emph{et~al.}
\newblock \bibinfo{title}{Kinetic nature of grain boundary formation in
  as-grown {M}o{S}$_2$ monolayers}.
\newblock \emph{\bibinfo{journal}{Adv. Mater.}} \textbf{\bibinfo{volume}{27}},
  \bibinfo{pages}{4069--4074} (\bibinfo{year}{2015}).

\bibitem{chae2012}
\bibinfo{author}{Chae, S.} \emph{et~al.}
\newblock \bibinfo{title}{Direct observation of the proliferation of
  ferroelectric loop domains and vortex-antivortex pairs}.
\newblock \emph{\bibinfo{journal}{Phys. Rev. Lett.}}
  \textbf{\bibinfo{volume}{108}}, \bibinfo{pages}{167603}
  (\bibinfo{year}{2012}).

\bibitem{jungk2010}
\bibinfo{author}{Jungk, T.}, \bibinfo{author}{Hoffmann, {\'A}.},
  \bibinfo{author}{Fiebig, M.} \& \bibinfo{author}{Soergel, E.}
\newblock \bibinfo{title}{Electrostatic topology of ferroelectric domains in
  {YM}n{O}$_3$}.
\newblock \emph{\bibinfo{journal}{Appl. Phys. Lett.}}
  \textbf{\bibinfo{volume}{97}}, \bibinfo{pages}{012904}
  (\bibinfo{year}{2010}).

\bibitem{park2017dark}
\bibinfo{author}{Park, K.-D.}, \bibinfo{author}{Jiang, T.},
  \bibinfo{author}{Clark, G.}, \bibinfo{author}{Xu, X.} \&
  \bibinfo{author}{Raschke, M.~B.}
\newblock \bibinfo{title}{Radiative control of dark excitons at room
  temperature by nano-optical antenna-tip induced {P}urcell effect}.
\newblock \emph{\bibinfo{journal}{preprint arXiv:1706.09085, Nat. Nanotech.}}
  (\bibinfo{year}{2017}).

\bibitem{jin2016}
\bibinfo{author}{Jin, C.} \emph{et~al.}
\newblock \bibinfo{title}{Interlayer electron-phonon coupling in
  {WS}e$_2$/h{BN} heterostructures}.
\newblock \emph{\bibinfo{journal}{Nat. Phys.}} \textbf{\bibinfo{volume}{13}},
  \bibinfo{pages}{127--131} (\bibinfo{year}{2017}).

\bibitem{chikkaraddy2016}
\bibinfo{author}{Chikkaraddy, R.} \emph{et~al.}
\newblock \bibinfo{title}{Single-molecule strong coupling at room temperature
  in plasmonic nanocavities}.
\newblock \emph{\bibinfo{journal}{Nature}} \textbf{\bibinfo{volume}{535}},
  \bibinfo{pages}{127--130} (\bibinfo{year}{2016}).

\end{thebibliography}

\vskip 1cm
\noindent
{\bf Acknowledgements}

\noindent
The authors would like to thank Joanna M. Atkin for insightful discussions.
We thank Xiaobo Yin and Manfred Fiebig for providing the MoS$_2$ sample and the YMnO$_3$ sample, respectively.
We acknowledge funding from the U.S. Department of Energy, Office of Basic Sciences, Division of Material Sciences and Engineering, under Award No. DE-SC0008807.
We also acknowledge support provided by the Center for Experiments on Quantum Materials (CEQM) of the University of Colorado.
\\

\noindent
{\bf Competing financial interests}

\noindent
The authors declare no competing financial interests.

\end{document}